\documentclass[aoas,MSNbibl,nameyear,dvips]{arximspdf}
\usepackage{dcolumn,breakurl}
\usepackage{graphicx}

\doi{10.1214/13-AOAS665} %kopijuoti is PTS
\volume{7}
\issue{4}
\pubyear{2013}
\firstpage{2062}
\lastpage{2080}

\makeatletter

\let\sv@tabnotetext\tabnotetext
\let\sv@tabnotemark@fmt\tabnotemark@fmt
\long\def\legend#1{{\let\tabnote@indent\leavevmode\sv@tabnotetext[]{}{#1}}}

\newcolumntype{d}[1]{D{.}{.}{#1}}

\makeatother

\begin{document}
\begin{frontmatter}

\title{Evaluating costs with unmeasured confounding: A~sensitivity
analysis for the treatment effect}
\runtitle{Sensitivity analysis for costs}

\begin{aug}
\author[A]{\fnms{Elizabeth A.} \snm{Handorf}\corref{}\thanksref{t1,m1}\ead[label=e1]{elizabeth.handorf@fccc.edu}},
\author[B]{\fnms{Justin E.} \snm{Bekelman}\thanksref{t2,m2}\ead[label=e2]{bekelman@uphs.upenn.edu}},\\
\author[C]{\fnms{Daniel F.} \snm{Heitjan}\thanksref{t3,m2}\ead[label=e3]{dheitjan@upenn.edu}}
\and
\author[C]{\fnms{Nandita} \snm{Mitra}\thanksref{t3,m2}\ead[label=e4]{nanditam@upenn.edu}}
\runauthor{Handorf, Bekelman, Heitjan and Mitra}
\affiliation{Fox Chase Cancer Center\thanksmark{m1} and University of
Pennsylvania\thanksmark{m2}}
\address[A]{E. A. Handorf\\
Biostatistics and Bioinformatics Facility \\
Fox Chase Cancer Center \\
Temple University Health System \\
333 Cottman Avenue\\
Philadelphia, Pennsylvania 19111\\
USA\\
\printead{e1}}
\address[B]{J. E. Bekelman\\
Department of Radiation Oncology\\
Perelman School of Medicine\\
University of Pennsylvania \\
3400 Spruce Street\\
Philadelphia, Pennsylvania 19104 \\
USA\\
\printead{e2}}
\address[C]{D. F. Heitjan\\
N. Mitra \\
Department of Biostatistics and Epidemiology \\
Perelman School of Medicine \\
University of Pennsylvania \\
423 Guardian Drive \\
Philadelphia, Pennsylvania 19104-6021\\
USA\\
\printead{e3}\\
\hphantom{E-mail: }\printead*{e4}}
\end{aug}

\thankstext{t1}{Supported in part by the NCI Grant UC2CA148310, and NIH Grant P30 CA 06927.}
\thankstext{t2}{Supported in part by the Thomas B. and Jeannette E. Laws McCabe Fund, the Abramson Cancer Center core grant (USPHS Grant P30 CA 016520), and NCI K07 CA163616.}
\thankstext{t3}{Supported in part by The Abramson Cancer Center core grant (USPHS Grant P30 CA 016520).}

% HISTORY:
\received{\smonth{8} \syear{2012}}
\revised{\smonth{6} \syear{2013}}

% ABSTRACT
%
\begin{abstract}
Estimates of the effects of treatment on cost from observational
studies are subject to bias if there are unmeasured confounders. It is
therefore advisable in practice to assess the potential magnitude of
such biases. We derive a general adjustment formula for loglinear
models of mean cost and explore special cases under plausible
assumptions about the distribution of the unmeasured confounder. We
assess the performance of the adjustment by simulation, in particular,
examining robustness to a key assumption of conditional independence
between the unmeasured and measured covariates given the treatment
indicator. We apply our method to SEER-Medicare cost data for a stage
II/III muscle-invasive bladder cancer cohort. We evaluate the costs
for radical cystectomy \textit{vs.} combined radiation/chemotherapy, and
find that the significance of the treatment effect is sensitive to
plausible unmeasured Bernoulli, Poisson and Gamma confounders.
\end{abstract}

% KEYWORDS

\end{frontmatter}

%s1 #&#
\section{Introduction}\label{sec1}

Payers and health care providers, such as Medicare, private insurers
and hospitals, routinely collect data on medical expenditures
[\citet
{Hornberger1997}]. Some of these data sources are readily available and
have high external validity [\citet{Black1996}] and therefore are widely
used to compare treatment costs. Like all observational data, however,
they are subject to confounding. That is, if all relevant confounders
are measured and present in the data set, we can find a consistent
estimate of the treatment effect by using a correctly specified
statistical model for cost. If some confounders are unmeasured or
unavailable, the model-based estimate is potentially biased. Here we
develop a method to assess the sensitivity to potential unmeasured
confounders when evaluating the effect of a medical intervention on
mean cost.

Cost data are nonnegative and often highly skewed, features that can
be readily described statistically in the generalized linear model
framework. For example, a Gamma regression with a log link is often a
suitable model for cost data [\citet{Dodd2006}]. The situation is more
complex when some costs are censored. Costs accrue over time, and they
may accrue at different rates among individuals. Therefore, cost at
the time of censoring is generally correlated with cost at the time of
ultimate failure, making censoring informative [\citet
{Lin1997,Etzioni1999}]. One popular method which accounts for this is Inverse
Probability Weighting (IPW), where observed costs are weighted by the
inverse probability of censoring [Lin (\citeyear{Lin2000,Lin2003}),
Bang and Tsiatis (\citeyear{Bang2000,Bang2002}),
\citet{Willan2002,Tian2007}]. In the
context of observational cost
data, the method of \citet{Lin2003} is particularly useful. He applies
IPW to generalized linear models, allowing adjustment for confounders
in a Gamma regression. Other approaches which handle informative
censoring are based on Bayesian methods or use multi-part models
[\citet{Heitjan2004,Basu2010,Liu2008,Liu2009}].

The methods cited above do not address the potential bias due to
unmeasured confounding. A natural approach to evaluate this bias is a
sensitivity analysis, in which one posits models for the distribution
of an unmeasured confounder and its effects on cost, then evaluates
their effects on estimates of the treatment effect. Researchers have
developed various methods implementing this idea
[\citet
{Rosenbaum1983,Schlesselman1978,Yanagawa1984,Axelson1988,Gail1988,Flanders1990,McCandless2007,Hosman2010}].
Further developments include the method of \citet{Lin1998}, in
which the
authors present a formula to determine the magnitude of the bias due to
an unmeasured Bernoulli or Normal covariate. \citet{Mitra2007} extended
this idea to assess the sensitivity of survival outcomes using a
Weibull model.

As we have indicated, the method of \citet{Lin1998} assesses the
sensitivity of the treatment effect to unknown binomial or normal
confounders. \citet{wang2006} have shown that in a matched pairs study,
conclusions are most sensitive to a Bernoulli confounder and, thus,
such an analysis is sufficient to describe maximal sensitivity. It is
unclear whether an analogous result holds in cohort studies, however,
or indeed whether the exploration of maximal sensitivity is most
desirable in every context. Therefore, in this paper we generalize the
Lin method to derive corrections for a broader range of distributions
of unmeasured confounders. For example, applying our formula for a
Gamma confounder, one can adjust for an unmeasured skewed
continuous\vadjust{\goodbreak}
variable such as personal income. Or using our formula for a Poisson
confounder, one can adjust for an unmeasured count variable such as
pack-years smoked.

A major drawback of the Lin method and related sensitivity analyses for
the effects of an unmeasured confounder is the assumption of
independence between measured and unmeasured confounders given
treatment status. \citet{Hernan1999} demonstrated that such an
assumption is implausible, as conditioning on treatment induces a
correlation even between unrelated covariates. \citet{Vanderweele2008}
showed how one of Lin's formulas can apply in certain circumstances
with relaxed assumptions, and \citet{Vanderweele2011} developed bias
adjustment formulas that are valid more generally without requiring
conditional independence. The implementation of these approaches can be
complex, however, and a simpler alternative such as the Lin method
would be invaluable in applications. In this paper, therefore, we both
extend the basic Lin method to a range of loglinear models and use
simulations to explore the limits of its validity under departures from
conditional independence.

We demonstrate our method with a comparison of mean medical costs for
two treatments for stage II/III muscle-invasive bladder cancer. As
this is the ninth costliest cancer [\citet{NCIcostcancer}], it is of
substantial interest to determine cost-effective treatments for it.
Here, we compare the lifetime costs for radical cystectomy, an
aggressive surgical procedure that is the current standard of care, to
combined radiotherapy and chemotherapy, an alternative organ-sparing
curative treatment. The data are censored because many registry
participants were still living when the database was closed. We base
our cost comparisons on observational data from the linked
SEER-Medicare registry, assessing the sensitivity of the treatment
effect to potential Bernoulli, Poisson and Gamma unmeasured
confounders.

The paper is organized as follows: in Section~\ref{sec2} we develop a general
formula to quantify the magnitude of the bias of treatment effect due
to an unknown confounder, and derive the correction for Poisson and
Gamma confounders. We further show how these formulas and those of Lin
et al. [\citet{Lin1998}] can be used to assess sensitivity with cost
data, including censored costs. We specifically address criticisms that
have been raised of the assumption of conditional independence of
observed and unobserved covariates given treatment. In Section~\ref{sec3} we
assess the performance of our method using Monte Carlo simulations,
evaluating robustness to departures from conditional independence. In
Section~\ref{sec4} we apply our method to the SEER bladder cancer data.

%s2 #&#
\section{Bias with an unmeasured confounder}\label{sec2}

We first consider uncensored costs. Assuming a generalized linear model
(GLM) with a log link, we define the true model as
\[
\mathrm{E}(Y | X, Z, U) = \exp\bigl( \alpha+ \beta_0 X+
\phi_0 U + \psi_1 X U+ \theta^{\prime} Z\bigr),
\]
where $Y$ is the total cost, $X$ is the treatment of interest ($X=0$
for control and $X=1$ for treated), $U$ is the unmeasured confounder
and $Z$ is a vector of measured covariates. The true regression
parameters are $\alpha$, $\beta_{0}$, $\theta$, $\phi_{0}$ and
$\psi
_{1}$. We allow the unmeasured confounder $U$ to have different effects
on cost in the treatment groups, so $\beta_0$ is the effect of $X$ when
$U=0$, and $\beta_0+\psi_{1}$ is the effect of $X$ when $U$ is nonzero.
Likewise, $\phi_{0}$ is the effect of $U$ in the control group, and
$\phi_0+\psi_{1}$ is the effect of $U$ in the treated group. For
convenience, we express the model as
%
%e1 #&#
%
\begin{equation}\label{eqtruemodel}
\mathrm{E}(Y | X, Z, U) = \exp\bigl( \alpha+ \beta_0 X+
\gamma_X U + \theta^{\prime} Z\bigr).
\end{equation}
This is equivalent to the model specified above, where $\gamma_0$ is
the effect of $\mathit{U}$ in the control group and $\gamma_1$ is the
effect of $\mathit{U}$ in the treated group. For simplicity, we
henceforth suppress the subscript in $\beta_0$.

In practice, we cannot fit model (\ref{eqtruemodel}) because we do not
know the value of $\mathit{U}$. We define the reduced model, which we
can fit, as
%
%e2 #&#
%
\begin{equation}\label{eqapparentmodel}
\mathrm{E} (Y | X, Z) = \exp\bigl( \alpha^* + \beta^* X+ \theta^{*\prime} Z
\bigr).
\end{equation}
Fitting model (\ref{eqapparentmodel}) yields an estimate of the
apparent treatment effect $\beta^*$.

By conditional expectation,
\begin{eqnarray*}
\mathrm{E}(Y | X, Z) &=& \int_{-\infty}^\infty\mathrm
{E}(Y | X, Z, U)\,dF(U | X, Z )
\\
&=& \exp\bigl(\alpha+ \beta X + \theta^{\prime} Z\bigr) \int
_{-\infty}^\infty\exp(\gamma_X U)\,dF(U | X,
Z)
\\
&=& \exp\bigl(\alpha+ \beta X + \theta^{\prime} Z\bigr)
\mathrm{M}_{U |
X,Z}(\gamma_X),
\end{eqnarray*}
where $\mathrm{M}_{U | X,Z}(\gamma_X)$ is the moment generating
function (m.g.f.) of $U|X,Z$. If we assume that $U$ is conditionally
independent of $Z$ given $X$ so that $\mathrm{M}_{U | X,Z}(\gamma_X) =
\mathrm{M}_{U | X}(\gamma_X)$, the following simplification is possible:
% % % % % % % % % % % % % % % FIX!!!!
%
%e3 #&#
%
\begin{eqnarray}\label{eqfollowingsimp}
\mathrm{E}(Y | X, Z) &=& \exp\bigl( \alpha+ \ln\bigl(\mathrm{M}_{U | 0}(
\gamma_0)\bigr)
\nonumber\\[-8pt]\\[-8pt]
&&{} + X \bigl[\beta+ \ln\bigl(\mathrm{M}_{U | 1}(\gamma_1)
\bigr) -\ln\bigl(\mathrm{M}_{U | 0}(\gamma_0)\bigr)\bigr] +
\theta^{\prime} Z\bigr).
\nonumber
\end{eqnarray}
Setting (\ref{eqapparentmodel}) equal to (\ref{eqfollowingsimp}) gives
%
%e4 #&#
%
\begin{equation}\label{eqmain}
\beta= \beta^* - \ln\bigl(\mathrm{M}_{U | X=1}(\gamma_1)\bigr)
+ \ln\bigl(\mathrm{M}_{U | X=0}(\gamma_0)\bigr).
\end{equation}
Thus, we can conduct a sensitivity analysis by positing distributions
for $U|X$ and values for $\gamma_X$ in the treated and control groups.
Equation (\ref{eqmain})
holds for any distribution of $U$ that one can characterize by a
moment-generating function, a large class of distributions. In the
following sections we show the solutions for $\beta$ when $U$ is
distributed as Poisson or Gamma. We also present the formulas developed
by \citet{Lin1998} for Bernoulli and Normal unmeasured
confounders, and
show how they are applicable in the cost setting.

%s2.1 #&#
\subsection{Poisson unmeasured confounder}\label{sec2.1}

For $U \sim \operatorname{Poisson}(\lambda_{X,Z})$, we have
\[
\mathrm{M}_{U | X,Z}(\gamma_X) = \exp\bigl(
\lambda_{X,Z}\bigl(e^{\gamma
_X}-1\bigr) \bigr),
\]
and under conditional independence ($\lambda_{X,Z}=\lambda_{X}$),
(\ref{eqmain}) becomes
\[
\beta= \beta^* - \bigl(\lambda_1 e^{\gamma_1} -
\lambda_0 e^{\gamma_0} + \lambda_0 -
\lambda_1\bigr),
\]
where $\lambda_0$ is the mean of $U$ in the control group and $\lambda
_1$ is the mean in the treated group, $\beta$ is the true treatment
effect, $\beta^*$ is the observed treatment effect, and $\gamma_0$ and
$\gamma_1$ are the effects of $U$ in the control and treated groups,
respectively.

%s2.2 #&#
\subsection{Gamma unmeasured confounder}\label{sec2.2}
%2.2 Gamma unmeasured confounder

An unobserved covariate
\[
U \sim\operatorname{Gamma}(\theta_{X,Z},\kappa_{X,Z})
\]
has m.g.f.
\[
\mathrm{M}_{U | X,Z}(\gamma_X) = (1-\theta_{X,Z}
\gamma_{X})^
{-\kappa_{X,Z}}.
\]
Under conditional independence ($\theta_{X,Z}=\theta
_{X}$, $\kappa_{X,Z}=\kappa
_{X}$), (\ref{eqmain}) implies
\[
\beta= \beta^* - \ln\frac{(1-\theta_0\gamma_0)^{\kappa
_0}}{(1-\theta
_1\gamma_1)^{\kappa_1}},
\]
where the mean and variance of $U$ are $\kappa_{0}\theta_0$ and
$\kappa
_{0}\theta_0^2$ for the control group and $\kappa_{1}\theta_1$ and
$\kappa_{1}\theta_1^2$ for the treated group.

%s2.3 #&#
\subsection{Bernoulli and Normal unmeasured confounders}\label{sec2.3}
%2.3 Bernoulli and Normal unmeasured confounders

Lin, Psaty and Kronmal
(\citeyear{Lin1998}) derived the relationship between the true and apparent
treatment effects for a binary outcome with a log link when unmeasured
confounders are Bernoulli or Normal. Because the derivation also
applies to a continuous outcome with a log link, their results are
special cases of (\ref{eqmain}). Specifically, if $U\sim
\operatorname{Bernoulli}(\pi_{X,Z})$, and $U$ is conditionally independent of $Z$
$(\pi_{X,Z}=\pi_{X})$, then
\[
\beta= \beta^* - \ln\frac{e^{\gamma_1}\pi_1 +(1-\pi
_1)}{e^{\gamma
_0}\pi_0 +(1-\pi_0)},
\]
where $\pi_0$ is the prevalence of $U$ in the control group, $\pi_1$ is
the prevalence of $U$ in the treated group, and the other parameters
are as above. Similarly, if $U\sim \operatorname{Normal}(\mu_{X,Z},1)$ and again is
conditionally independent of $Z$ ($\mu_{X,Z}=\mu_{X}$), then
\[
\beta= \beta^* + \bigl(\mu_0\gamma_0 -
\mu_1\gamma_1 + \bigl(\gamma_0^2
- \gamma_1^2\bigr)/2 \bigr),
\]
where $\mu_j$ is the mean of $U$ in treatment group $j$, $j=0,1$, and
$\beta$, $\beta^*$, $\gamma_0$ and $\gamma_1$ are as above.

%s2.4 #&#
\subsection{Assumption of conditional independence}\label{sec2.4}
%2.4 Assumption of conditional independence within treatment strata

The conditional independence assumption (that $U$ is conditionally
independent of $Z$ given $X$) cannot be tested in practice since $U$
is\vadjust{\goodbreak}
not observed. However, it has been well-established that the assumption
cannot actually hold in practice. Conditional independence would hold
if the marginal correlation perfectly cancels the conditional
correlation or if the covariates $Z$ are not truly confounders, but
either condition is highly improbable. \citet{Hernan1999} observed that
conditional independence cannot hold if $Z$ and $U$ are marginally
independent and are both confounders. \citet{Vanderweele2008} showed
that closed-form solutions are available for normal unmeasured
confounders and for a binary confounder with a normal outcome under a
relaxed additivity assumption. \citet{Vanderweele2011} developed general
formulas for calculating an adjusted estimate of the treatment effect
without assumptions about the relationship between the measured and
unmeasured covariates. In the most general case one must specify
$\mathrm{E}(Y)$ at each level of $X$, $U$ and $Z$. When $U$ or $Z$ is
continuous, this would be very difficult to implement in practice. For
the case where the relationship between $Y$ and $U$ is constant across
levels of $Z$, as would occur in a linear model, \citet{Vanderweele2011}
developed simplified formulas that require fewer sensitivity
parameters. In our loglinear regression setting, however, an additional
bootstrapping approach would be required to use these formulas.

Although we do not expect our corrections to hold exactly in practice
due to weak tenability of the conditional independence assumption, if
conditional dependence is modest, they may still be useful. In Section~\ref{sec3.2} we present simulations to investigate the effect of dependence
within treatment strata.

%s2.5 #&#
\subsection{Analysis with censored costs}\label{sec2.5}
%2.5 Adjustment for unmeasured confounder with censored costs

We derived our adjustment formulas assuming uncensored data;
nevertheless, because they are statements about parameters, they are
valid also in the censored case provided that one estimates the
parameters consistently. \citet{Lin2003} developed a generalized
regression method, \mbox{appropriate} for censored costs, that combines the
generalized estimating equation approach with inverse probability of
censoring weighting. This approach produces consistent estimates but
forfeits efficiency because it uses only the uncensored data.
\citet
{Bang2002} developed a more efficient estimator for linear regression
that uses all of the data, but an extension to the GLM has yet to be
developed. We therefore recommend using the \citet{Lin2003} approach,
but note that our formulas are applicable with any method that gives
consistent estimates for a loglinear mean model.

%s2.6 #&#
\subsection{Sensitivity analysis for cost regression}\label{sec2.6}
%2.6 Sensitivity analysis for cost regression

The sensitivity analysis proceeds in four steps: first, estimate $\beta
^*$ in (\ref{eqapparentmodel}) by any consistent method. Second,
specify the type of the unmeasured confounder: for example, Poisson for
a count confounder; Gamma for skewed, positive continuous; Normal for
symmetric continuous; and Bernoulli for binary. Third, hypothesize
possible parameter values $\eta$ for the distributions of the
unmeasured confounder in the treated and control groups, and also the
possible effect $\gamma$ of the unmeasured confounder on cost. Finally,
apply the correction via (\ref{eqmain}) to obtain $\hat{\beta}$, the
adjusted treatment effect.

The analysis assumes that the sensitivity parameters $\eta$ are known.
Because the adjustment is in every case additive, it follows that
\[
\operatorname{Var}(\hat{\beta} | X, Z, \eta) = \operatorname{Var}\bigl
(\hat{\beta}^*|
X, Z
\bigr).
\]
We use this relationship to calculate confidence intervals and
determine the significance of the adjusted treatment effect, as was
done by \citet{Lin1998,Mitra2007} and
\citet{Vanderweele2011}. Note that this does not imply that
$\operatorname{Var}(\hat{\beta} | X, Z,U) = \operatorname{Var}
(\hat{\beta}^*| X,Z)$. If we were
able to measure~$U$, the variance of our estimate of $\beta$ would be
smaller, however, our correction does not reduce the uncertainty of our
estimate, only the bias. Therefore, the appropriate variance estimate
for the corrected $\hat{\beta}$ is $\operatorname{Var}(\hat{\beta}^*| X,Z)$. We
confirm this \textit{via} simulation in Section~\ref{sec3.1} where we
show that coverage probabilities are close to the nominal level,
indicating that $\operatorname{Var}(\hat{\beta}^*| X,Z)$ is a reasonable estimate
for the variance of the corrected treatment effect.

The investigator should repeat this procedure for a range of
combinations of parameter values and effect sizes, assessing the
plausibility of any combination of inputs that materially changes the
conclusions. If the significance of the treatment effect changes due to
unmeasured confounders with a small effect on cost and similar
distributions in the treated and control groups, one can conclude that
inferences are vulnerable to the effects of unmeasured confounding. We
demonstrate the procedure in Section~\ref{sec4}.

%s3 #&#
\section{Simulations}\label{sec3}

The general plan of the simulations was to generate costs knowing the
values of both measured and unmeasured covariates, then fit the reduced
regression omitting the unmeasured covariate and compute the adjusted
treatment effect. We evaluated performance by the bias of the adjusted
estimate and the coverage probability of the resulting 95\% confidence interval.

%s3.1 #&#
\subsection{Method performance}\label{sec3.1}

Initial simulations evaluated the performance of the method under
independence of the unmeasured confounder $U$ and measured confounder
$Z$ given treatment status $X$. We constructed 1000 replications in
each simulation, with 100, 200 and 500 subjects per treatment stratum.
We drew $U$ independently from its distribution $f_U(U | X, \eta_X)$
using the parameter values $\eta_X$ shown in Table~\ref{tabl1}. For example, a
%
%t1 #&#
%
\begin{table}
\caption{Monte Carlo simulations of the adjusted
treatment effect, $n=100$ per treatment stratum}
\label{tabl1}
\begin{tabular*}{\tablewidth}{@{\extracolsep{\fill}}l l c c c c@{}}
\hline
& & \multicolumn{4}{c@{}}{\textbf{Mean of} $\bolds{\hat{\beta}}$\tabnoteref{ta}
\textbf{(coverage probability)}} \\ [-4pt]
& & \multicolumn{4}{c@{}}{\hrulefill}
\\
\textbf{Distribution} & \multicolumn{1}{c}{$\bolds{\gamma}$} & \textbf{Uncensored}
& \textbf{25\% censoring} & \textbf{50\% censoring}
& \multicolumn{1}{c@{}}{\textbf{75\% censoring}}\\
\hline
Bernoulli\tabnoteref{tb} & 0.25 & 0.999 (0.963) & 1.001 (0.925) & 1.001
(0.933) & 0.997 (0.817) \\
& 0.5 & 1.001 (0.958) & 0.998 (0.936) & 1.003 (0.885) & 1.004 (0.790) \\
& 0.75 & 1.000 (0.954) & 1.004 (0.924) & 0.999 (0.918) & 1.005 (0.780)
\\
& 1 & 1.003 (0.946) & 1.007 (0.935) & 1.008 (0.909) & 1.016 (0.744) \\
Normal\tabnoteref{tc} & 0.25 & 1.001 (0.931) & 0.996 (0.950) & 1.000 (0.911) &
1.000 (0.856) \\
& 0.5 & 1.002 (0.949) & 1.000 (0.950) & 1.006 (0.910) & 0.995 (0.841) \\
& 0.75 & 1.010 (0.946) & 0.994 (0.928) & 1.003 (0.924) & 0.983 (0.841)
\\
& 1 & 1.006 (0.941) & 0.995 (0.923) & 0.982 (0.892) & 1.028 (0.815) \\
Poisson\tabnoteref{td} & 0.25 & 1.002 (0.959) & 1.003 (0.943) & 1.001
(0.935) & 0.998 (0.841) \\
& 0.5 & 0.994 (0.939) & 0.995 (0.936) & 0.986 (0.899) & 0.994 (0.804) \\
& 0.75 & 0.981 (0.924) & 0.979 (0.892) & 0.965 (0.873) & 0.895 (0.768)
\\
& 1 & 0.925 (0.894) & 0.922 (0.850) & 0.877 (0.828) & 0.811 (0.758) \\
Gamma\tabnoteref{te} & 0.25 & 1.001 (0.957) & 0.999 (0.948) & 1.002
(0.924) & 0.996 (0.855) \\
& 0.5 & 0.999 (0.947) & 0.997 (0.919) & 0.996 (0.910) & 0.997 (0.851) \\
& 0.75 & 0.998 (0.951) & 0.989 (0.913) & 0.970 (0.893) & 0.957 (0.840)
\\
& 1 & 0.931 (0.881) & 0.908 (0.853) & 0.915 (0.847) & 0.882 (0.780) \\
\hline
\end{tabular*}
\tabnotetext[\mbox{$\dagger$}]{ta}{For all models, the true value of
$\beta$
is 1.}
\tabnotetext[\mbox{$\diamond$}]{tb}{Unmeasured Bernoulli covariate with
$\pi
_0=0.3,\pi_1=0.866$.}
\tabnotetext[\mbox{$\ast$}]{tc}{Unmeasured Normal covariate with $\mu
_0=0,\mu
_1=1,\sigma_0=\sigma_1=1$.}
\tabnotetext[\mbox{$\ddagger$}]{td}{Unmeasured Poisson covariate with
$\lambda_0=1,\lambda_1=1.58$.}
\tabnotetext[\mbox{$\bigtriangleup$}]{te}{Unmeasured Gamma covariate with
$\kappa_0=\kappa_1=0.75, \theta_0=0.5,\theta_1=0.868$.}
\end{table}
Poisson unmeasured confounder was $\operatorname{Poisson}(\lambda=1)$ in the control
group and $\operatorname{Poisson}(\lambda=1.58)$ in the treated group. We next drew a
scalar measured confounder $Z$, which can represent either a single
confounder or a linear combination of a vector $Z$. Each $Z$ was an
independent draw from a Normal distribution with variance 1 and mean
$0$ in the control group and mean $1$ in the treated group. We computed
the mean cost according to (\ref{eqtruemodel}), where $\alpha=5$,
$\beta=1$, $\theta=1$, and $\gamma_0=\gamma_1=\gamma$. We drew cost
outcomes from $f_Y(Y | X, Z, U)$, where $f_Y$ was the Gamma with
variance equal to the mean. We drew censoring indicators using a
Bernoulli with varying probabilities as indicated in Table~\ref{tabl1}. We drew
failure times from the Exponential with mean 5 and censoring times
uniformly over $[0,10]$, independently of both treatment group and costs.
It was only necessary to identify the subjects with censored costs, not
generate their actual censored cost values, because the IPW method uses
only noncensored observations [\citet{Lin2003}]. In each
replication, we
calculated $\hat{\beta}^*$ from the apparent model and $\hat{\beta}$
from the adjustment formula. For uncensored data, we used the Gamma
GLM, and for censored data the IPW estimator.

Results of the simulation with 100 subjects per treatment stratum
appear in Table~\ref{tabl1}. We observe that the correction worked well for
Bernoulli and Normal confounders with modest censoring. As the
proportion censored increased, coverage probability declined from the
nominal level, although the bias remained small. For example, the
treatment effect with a Bernoulli unmeasured confounder of effect size
$\gamma= 0.5$ had a coverage probability of 0.96 with $0\%$ censoring
and 0.79 with $75\%$ censoring, yet the mean of $\hat{\beta}$ remained
close to its true value of 1 regardless of censoring. The method worked
well for Poisson and Gamma confounders with moderate effect sizes
($\gamma\leq0.5$), but as the effect of $U$ drew close to the effect
of~$X$, the corrected $\hat{\beta}$ incurred a bias and coverage
probabilities fell below nominal levels. For example, with Gamma $U$
and uncensored costs, if the effect of $U$ was small $(\gamma= 0.25)$,
the bias of $\hat{\beta}$ was $0\%$ with a coverage probability of
0.96, but if the effect was large $(\gamma= 1)$, the bias was $-7\%$
with a coverage probability of 0.88. Increased proportions of censoring
further degraded the performance of the estimator. In the uncensored
case, increasing the sample size to 2000 per treatment stratum
resulted in negligible bias even for large effects (not shown). We
conclude that with substantial confounding the adjusted estimates may
incur bias in samples of moderate size.

With Gamma and Poisson confounders, 1\%--4\% of the reduced regression
models failed to converge. This occurred only when $U$ had a large
effect on cost ($\gamma\ge0.75$). Analysis of the simulation data
revealed that all true models converged, and reduced models failed only
when there were single large outliers. Therefore, divergence of the
reduced model may itself serve as an indicator of misspecification due
to an unmeasured confounder.

%s3.2 #&#
\subsection{Sensitivity to departures from conditional
independence}\label{sec3.2} A second set of simulations assessed
robustness to violations of conditional independence. Following
\citet{Hernan1999} and \citet {Vanderweele2008}, we generated
the treatment $X$ conditional on the confounders. We first drew
$Z\sim\operatorname{Normal}(1,1)$, then $U\sim f_U(U | Z=z)$ defined as
follows:
$\operatorname{Bernoulli}(\pi=\operatorname{expit}(0.5+0.2z))$;
$\operatorname{Normal}(\mu=1+ 0.1z,\sigma=1)$;
$\operatorname{Poisson}(\lambda=0.9+ 0.1z)$; or
$\operatorname{Gamma}(\kappa=0.5,\theta=0.65+0.2|z|)$. With these parameters the
unconditional correlation between $Z$ and $U$ is approximately $0.1$.
Next, we generated $X\sim\operatorname{Bernoulli}(\Pr(\mathit{X}=1)=
\operatorname{expit}(\phi_1+ \phi_2 \mathit{z}+ \phi_3\mathit{u}))$,
varying $\phi _1,\phi_2,\phi_3$ to obtain different patterns of
correlation. As before, we drew costs from a Gamma with mean function
(\ref{eqtruemodel}).

%
%f1 #&#
%
\begin{figure}

\includegraphics{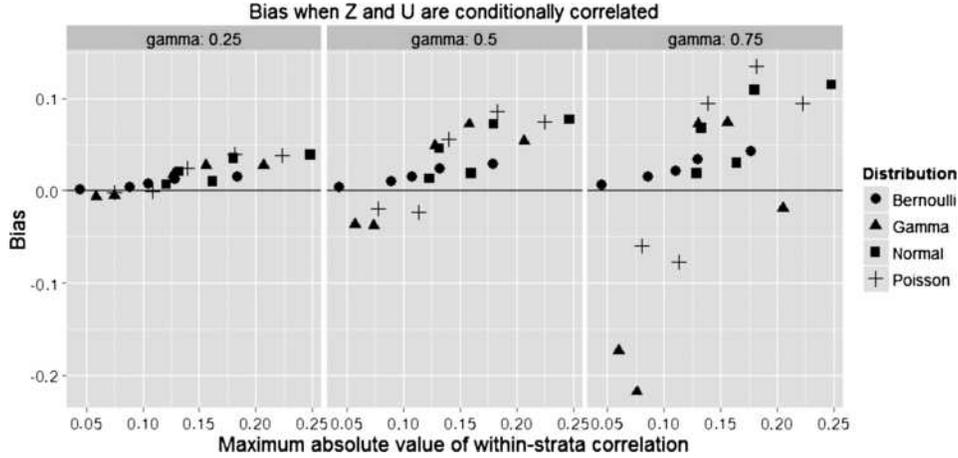}

\caption{Bias of treatment effect.}\label{figur1}
\end{figure}

\begin{figure}[b]

\includegraphics{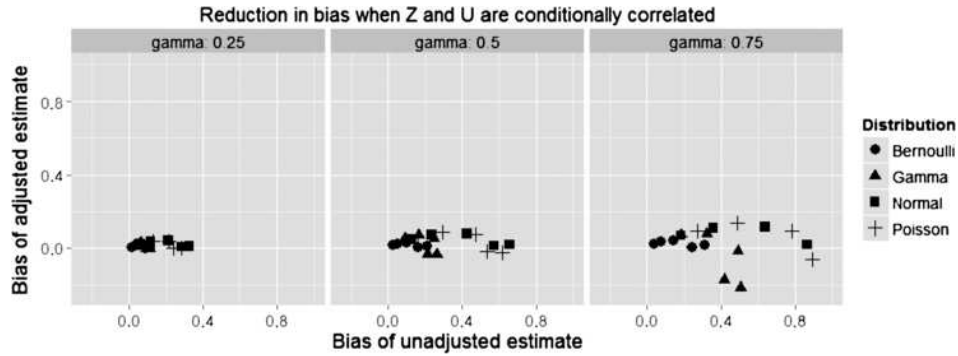}

\caption{Reduction in bias from unadjusted estimate.}\label{figur2}
\end{figure}

\begin{figure}

\includegraphics{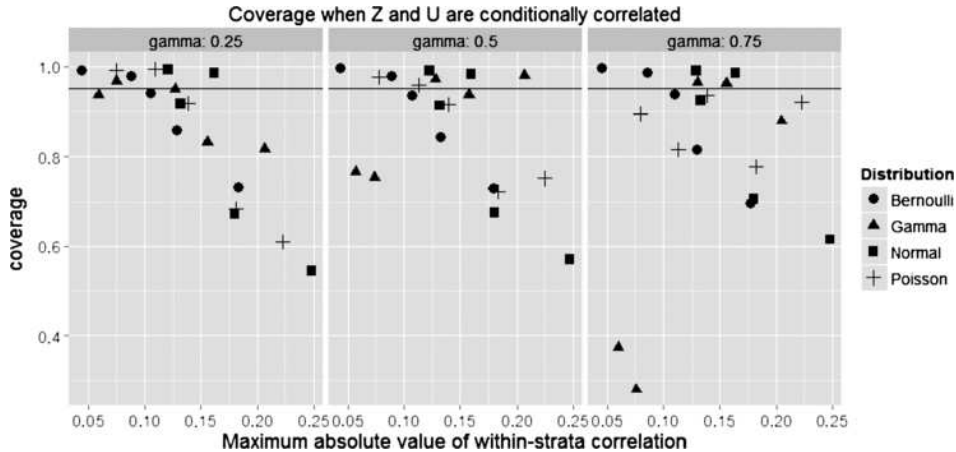}

\caption{Coverage of treatment effect.}\label{figur3}
\end{figure}

Figure~\ref{figur1} plots the bias in the corrected estimate vs. the maximum
absolute within-stratum correlation (the larger of the correlations in
the two treatment strata) for $n=500$ under 25\% censoring. We see that
with correlations above $0.15$, biases exceeded 5\% even for moderate
effect sizes, and the bias increased with the strength of $U$. Figure~\ref{figur2}
plots the bias of the adjusted estimate vs. that of the unadjusted
%
%f2 #&#
%
estimate, revealing that the adjustment was substantially more accurate
in all cases. Figure~\ref{figur3} plots the coverage probability vs.
the maximum
absolute within-stratum correlation. We note that effects on coverage
probability can be large for modest partial correlations, even when the
confounder effect size is also modest.

%
%f3 #&#
%

For example, consider a Bernoulli $U$ where $\gamma=0.75$ with $\phi
_1=-1$, $\phi_2=1$, and $\phi_3=2$. For these parameters, the partial
correlation of $U$ and $Z$ is $-0.086$ in the $X=1$ stratum and
$-0.046$ in the $X=0$ stratum. Whereas the unadjusted estimate of
$\beta
$ has a bias of 30.8\% with 0\% coverage, the adjusted treatment effect
has a bias of only 1.6\% and a coverage of 98.6\%.

In further simulations, we explored the relative importance of
correlations between $U$ and individual elements of $Z$ vs. the
aggregate correlation of $U$ with a summary of $Z$. We generated a
four-variate normal vector $(U,Z)$ where $U$ was a scalar unmeasured
confounder and $Z$ was a trivariate measured confounder, assuming two
different correlation models: in model 1, the correlations between $U$
and elements of $Z$ were all set to 0.1; in model 2, the correlations
were 0.3, $-0.4$ and 0. We then generated the treatment indicators and
computed the propensity scores [estimated $e = \operatorname
{Pr}(X=1|Z)$] [\citet
{Rosenbaum1983a}] to summarize the effects of $Z$. Although individual
partial correlations were larger in model~2 than model 1,
within-stratum correlations between $U$ and the propensity score were
moderate under both models (e.g., $-0.02$ in the $X=1$ stratum and
$-0.10$ in the $X=0$ stratum under model 2). Biases were 0\% for model
1 and 2\% for model 2. Thus, the simulation, although not exhaustive,
suggested that the total correlation of $U$ with the propensity score
is more important than individual correlations between $U$ and elements
of $Z$. We also considered the effect of the unconditional correlation
between $U$ and $Z$ on bias and coverage. Briefly, larger unconditional
correlations of 0.2 and 0.3 led to larger within-stratum correlations,
but the bias was largely driven by the within-stratum correlation.

Our simulations suggest that our correction is useful for moderately
strong unmeasured confounders when the total within-stratum correlation
between the unmeasured and measured confounders is less than 0.15.
While the adjusted treatment effect may have some residual bias under
conditional dependence, it can still give the investigator a sense of
how strongly an unmeasured confounder might influence the treatment effect.

%s4 #&#
\section{Bladder cancer study}\label{sec4}

We have applied our method to data from a cohort of stage II/III
bladder cancer patients diagnosed between 1995 and 2005 and appearing
in the linked SEER-Medicare data set [\citet
{Schrag20051118,KRobinYabroff05072008}]. We restrict analysis to the
two main treatments: the standard therapy of radical cystectomy, and a
bladder-sparing regimen consisting of radiotherapy and cisplatin-based
chemotherapy (rad/chemo). We excluded patients who were diagnosed
before age 65, for whom bladder cancer was not the first primary
malignancy, who did not have continuous Medicare part A/B coverage or
who had coverage from an HMO during the treatment period, or for whom
key data such as histology were missing. Radiotherapy and chemotherapy
could be given up to 3 months apart, so in order to avoid survivorship
bias, we excluded patients who died before 90 days. The final cohort
included 1860 patients, of whom 77.4$\%$ were treated with cystectomy.

We extracted payment data for Medicare parts A and B from the Carrier
Claims file, the Outpatient file, and the Medicare Provider Analysis
and Review Record. These files contained payment information for
physicians, outpatient services from institutions such as hospitals and
health clinics, and inpatient services from hospitals and long-term
care facilities [\citet{SEERmedicare2010}]. All payments were adjusted
to year 2000 dollars using the Medicare Economic Index [\citet
{KRobinYabroff05072008,MedicareEconInd}]. This method allowed us to
compare total costs of care rather than costs due to treatment only.
Two patients had zero costs, suggesting some incompleteness in the
payment data; we corrected for this by adding half of the lowest
nonzero cost to all observations [\citet{Mosteller1977}].

We modeled costs using a GLM with a log link and a gamma variance
function. Because the observed total costs were subject to censoring,
we estimated the model by the IPW method [\citet{Lin2003}]. Our analysis
considered all health care costs, for which the censoring rate was a
moderate 39.0\%.

Using a Gamma GLM accounting for censoring, we estimated the effect of
rad/chemo on cost, adjusting for the available measured covariates
tumor grade, sex, ethnicity, marital status, age, median income of the
census tract, size of the metropolitan area, number of comorbidities,
year of diagnosis and SEER site. We found that rad/chemo lowers the
cost of treatment by a factor of 0.873 (95\% CI: 0.793, 0.960).

%
%t2 #&#
%
\begin{table}
\tabcolsep=0pt
\caption{Observed correlations in bladder cancer
study}
\label{tabl2}
\begin{tabular*}{\tablewidth}{@{\extracolsep{\fill}}l d{2.3} d{2.3} d{2.3} c d{2.3}d{2.3}@{}}
\hline
& \multicolumn{3}{c}{\textbf{Corr with propensity score}} & & \multicolumn
{2}{c@{}}{\textbf{Largest individual corr}} \\[-4pt]
& \multicolumn{3}{c}{\hrulefill} & & \multicolumn
{2}{c@{}}{\hrulefill} \\
& \multicolumn{1}{c}{\textbf{Unconditional}} & \multicolumn{1}{c}{\textbf{Rad/chemo}}
& \multicolumn{1}{c}{\textbf{Cystectomy}} & & \multicolumn{1}{c}{\textbf{Rad/chemo}}
& \multicolumn{1}{c@{}}{\textbf{Cystectomy}} \\
\hline
Grade & & & & & & \\
\quad1 & 0.004 & -0.004 & 0.000& & 0.150 & 0.082 \\
\quad3 & 0.006 & -0.085 & 0.024& & 0.153 & 0.124 \\
\quad4 & -0.007 & 0.076 & -0.013& & 0.183 & 0.136 \\
\quad5 & -0.006 & 0.049 & -0.051& & 0.153 & 0.061 \\
& & & & & & \\
Sex & -0.021 & -0.035 & -0.059& & -0.405 & 0.373 \\
& & & & & & \\
Race & & & & & & \\
\quad White & 0.140 & 0.111 & 0.148& & -0.189 & -0.235 \\
\quad Black & -0.075 & -0.070 & -0.084& & 0.235 & 0.177 \\
\quad Other & -0.125 & -0.099 & -0.125& & 0.167 & 0.381 \\
& & & & & & \\
Hispanic & -0.074 & -0.001 & -0.071& & 0.264 & 0.202 \\
& & & & & & \\
Marital status & & & & & & \\
\quad Married & -0.003 & -0.001 & 0.012& & 0.384 & 0.373 \\
\quad Unmarried & 0.014 & 0.007 & 0.003& & 0.405 & 0.351 \\
\quad Unknown & -0.032 & -0.017 & -0.044& & 0.153 & 0.322 \\
& & & & & & \\
Age & 0.134 & 0.115 & 0.061& & 0.199 & 0.162 \\
& & & & & & \\
Urban Category & & & & & & \\
\quad1 & -0.053 & -0.066 & -0.043& & -0.408 & -0.311 \\
\quad2 & 0.013 & 0.009 & 0.005& & 0.328 & 0.297 \\
\quad3 & 0.048 & 0.067 & 0.045& & 0.490 & 0.359 \\
& & & & & & \\
Comorbidities & & & & & & \\
\quad0 & -0.043 & -0.044 & -0.019& & 0.097 & -0.097 \\
\quad1 & -0.071 & -0.075 & -0.051& & 0.128 & 0.066 \\
\quad2 or more & 0.091 & 0.093 & 0.058& & 0.140 & 0.108 \\
\hline
\end{tabular*}
\end{table}

%
%t3 #&#
%
\begin{table}
\def\arraystretch{0.9}
\tablewidth=300pt
\caption{Sensitivity of the estimated bladder
cancer treatment cost ratio
to an unmeasured Bernoulli confounder}
\label{tabl3}
\begin{tabular*}{\tablewidth}{@{\extracolsep{\fill}}l c d{1.2} c c@{}}
\hline
$\bolds{\pi_0}$ & \multicolumn{1}{c}{$\bolds{\pi_1}$}
& \multicolumn{1}{c}{\textbf{Effect of confounder}}
& \multicolumn{1}{c}{\textbf{Cost ratio}} &
\multicolumn{1}{c@{}}{$\bolds{95\%}$ \textbf{CI}} \\
\hline
0.7 & 0.5 & 1.1 & 0.89 & (0.81, 0.98) \\
0.8 & 0.4 & 1.1 & 0.91 & \textbf{(0.82, 1.00)} \\
0.8 & 0.3 & 1.1 & 0.92 & \textbf{(0.83, 1.01)} \\
[4pt]
0.7 & 0.5 & 1.25 & 0.91 & \textbf{(0.83, 1.00)} \\
0.8 & 0.4 & 1.25 & 0.95 & \textbf{(0.87, 1.05)} \\
0.8 & 0.3 & 1.25 & 0.97 & \textbf{(0.89, 1.07)} \\
[4pt]
0.7 & 0.5 & 1.5 & 0.94 & \textbf{(0.86, 1.04)} \\
0.8 & 0.4 & 1.5 & 1.02 & \textbf{(0.93, 1.12)} \\
0.8 & 0.3 & 1.5 & 1.06 & \textbf{(0.97, 1.17)} \\
\hline
\end{tabular*}
\legend{Unadjusted cost ratio${}={}$0.87 (95$\%$ CI 0.79--0.96).
\\
\textbf{Boldface} confidence intervals denote changes
from the unadjusted inference.
\\
$\pi_0 = \mbox{prevalence}$ of unmeasured confounder in
cystectomy group.
\\
$\pi_1 = \mbox{prevalence}$ of unmeasured confounder in
rad/chemo group.}
\end{table}

%s4.1 #&#
\subsection{Potential magnitude of conditional correlation}\label{sec4.1}

To determine the plausible extent of violations of the assumption $Z
\perp U | X$, we calculated partial correlations between each of the
measured confounders, $Z_j$, and the combined effect of the other
covariates estimated by the propensity score excluding $Z_j$. Some
individual pairwise correlations are substantial, for example, the
correlation between sex and marital status among the treated is
$-0.405$. When considering the aggregate correlation with all other
confounders, however, we found that correlations were typically
smaller, often less than 0.1 (see Table~\ref{tabl2}). The highest combined within-stratum
correlations were with race, with the largest observed correlation
being 0.148. Although we can never be sure of the magnitude of
correlations between $U$ and $Z$, if aggregate correlations with the
unmeasured confounder are no more extreme than the highest observed
value, our procedure will yield valid results.

Because our unadjusted estimate was statistically significant, we were
primarily interested in assessing potential unmeasured confounders that
would displace the effect of treatment toward the null. To this end, we
posited three potential confounders: access to health care, manifested
as a Bernoulli variable; smoking history, manifested as Poisson; and
personal income, manifested as Gamma.\newpage

%s4.2 #&#
\subsection{Unmeasured Bernoulli confounder}\label{sec4.2}

Although this cohort had Medicare insurance that guaranteed coverage of
their care, other factors that are unavailable in SEER-Medicare such as
lack of transportation or low physician availability may have limited
access [\citet{Penchansky1981}]. For example, if rad/chemo
patients had
poorer access, their overall medical costs may also have been lower. We
performed a sensitivity analysis to determine how strong this
unmeasured confounder would need to be to explain the treatment effect
on costs. As access to health care is a difficult concept to quantify,
we categorized it as binary, with $U=1$ representing good access and
$U=0$ poor access.

Our sensitivity analysis (Table~\ref{tabl3}) varied both the effect of $U$ on
expected cost and the distributions of $U$ in the treatment strata. We
see that moderate imbalances in $U$ produced tangibly different
inferences from the original analysis, even when the effect of the
unmeasured confounder was small. The adjusted treatment effect was
nonsignificant when the prevalence of good access to care was $80\%$ in
the cystectomy group and $40\%$ in the rad/chemo group and the
confounder had an effect of 1.1 on the cost ratio. Alternatively, if
the effect of access on cost was 1.25, prevalences of 70\% in the
cystectomy group and 50\% in the rad/chemo group produced
nonsignificant confidence intervals. We conclude that although the
apparent treatment effect is significant, this could be easily
explained by an unmeasured Bernoulli confounder.

%s4.3 #&#
\subsection{Unmeasured Poisson confounder}\label{sec4.3}
%4.2 Unmeasured Poisson confounder

Smoking is a major risk factor for bladder cancer [\citet{Baris2009}]
and for co-morbid conditions such as lung and\vadjust{\goodbreak} artery disease
[\citet
{Freund1993}]. Treating these conditions is likely to increase costs.
It is common to quantify smoking history in pack-years, defined as the
number of packs smoked per day times the length of smoking in years;
because this is a discrete count variable, it is natural to model it as
Poisson [see \citet{Wang2008}]. Smoking intensity is related to medical costs; a decrease of
20 pack-years has been associated with 10\% lower costs [\citet
{Leigh2005}]. This means that we could expect costs to increase by a
factor of approximately 1.005 per additional pack-year smoked. We
estimate the mean number of pack-years in our bladder cancer population
to be between 15 and 30 [\citet{Baris2009}]. If exposure differs between
treatment strata, the treatment effect may be subject to unmeasured confounding.

%
%t4 #&#
%
\begin{table}
\def\arraystretch{0.9}
\tablewidth=300pt
\caption{Sensitivity of the estimated bladder
cancer treatment cost ratio
to an unmeasured Poisson confounder}
\label{tabl4}
\begin{tabular*}{\tablewidth}{@{\extracolsep{\fill}}l c d{1.3} c c@{}}
\hline
$\bolds{\lambda_0}$ & $\bolds{\lambda_1}$ & \multicolumn{1}{c}{\textbf{Effect of confounder}}
& \textbf{Cost ratio} & \multicolumn{1}{c@{}}{$\bolds{95\%}$ \textbf{CI}} \\
\hline
15 & 13 & 1.005 & 0.88 & (0.80, 0.97) \\
15 & 11 & 1.005 & 0.89 & (0.81, 0.98) \\
17 & 11 & 1.005 & 0.90 & (0.82, 0.99) \\
19 & 11 & 1.005 & 0.91 & \textbf{(0.83, 1.00)} \\
19 & 9 & 1.005 & 0.92 & \textbf{(0.83, 1.01)} \\
[4pt]
15 & 13 & 1.01 & 0.89 & (0.81, 0.98) \\
15 & 11 & 1.01 & 0.91 & \textbf{(0.83, 1.00)} \\
17 & 11 & 1.01 & 0.93 & \textbf{(0.84, 1.02)} \\
[4pt]
30 & 28 & 1.005 & 0.88 & (0.80, 0.97) \\
30 & 26 & 1.005 & 0.89 & (0.81, 0.98) \\
32 & 26 & 1.005 & 0.90 & (0.82, 0.99) \\
34 & 26 & 1.005 & 0.91 & \textbf{(0.83, 1.00)} \\
34 & 24 & 1.005 & 0.92 & \textbf{(0.83, 1.01)} \\
[4pt]
30 & 28 & 1.01 & 0.89 & (0.81, 0.98) \\
30 & 26 & 1.01 & 0.91 & \textbf{(0.83, 1.00)} \\
32 & 26 & 1.01 & 0.93 & \textbf{(0.84, 1.02)} \\
\hline
\end{tabular*}
\legend{Unadjusted cost $\mbox{ratio} = 0.87$ (95$\%$ CI 0.79--0.96).\\
\textbf{Boldface} confidence intervals denote changes
from the unadjusted inference.\\
$\lambda_0 = \mbox{mean}$ of unmeasured confounder in
cystectomy group.\\
$\lambda_1 = \mbox{mean}$ of unmeasured confounder in
rad/chemo group.}
\end{table}

Table~\ref{tabl4} presents the sensitivity analysis for smoking history. A
single pack-year of smoking will not affect costs substantially, so
moderate differences in smoking history between the groups are required
to change the significance of the treatment effect. If the effect of a
single pack-year on cost is 1.005, an average of 34 pack-years in the
cystectomy group and 26 pack-years in the rad/chemo group would yield a
nonsignificant confidence interval. Because a difference this large is
plausible, we deem that the treatment effect is sensitive to smoking
history manifested as a Poisson confounder.

%s4.4 #&#
\subsection{Unmeasured Gamma confounder}\label{sec4.4}
%4.3 Unmeasured Gamma confounder

Income is positively associated with health care spending at the
population level [\citet{Matteo2003,CMSMCBS}]. Median income for the
census tract was available in SEER-Medicare data, and our analysis
showed no evidence of imbalance on aggregate, but there was still a
possibility of imbalance at the patient level. We therefore
hypothesized that an effect of income on spending would be small, and a
potential imbalance of income between the patient groups would also be
small.

%
%t5 #&#
%
\begin{table}
\def\arraystretch{0.9}
\caption{Sensitivity of the estimated bladder cancer treatment cost
ratio to an unmeasured Gamma~confounder} \label{tabl5}
\begin{tabular*}{\tablewidth}{@{\extracolsep{\fill}}l c d{1.2} c c@{}}
\hline
\multicolumn{1}{@{}l}{\textbf{Mean}$\bolds{_0}$\textbf{$\bolds{/}$Mean}$\bolds{_1}$}
& \multicolumn{1}{c}{\textbf{Var}$\bolds{/}$\textbf{Mean}$\bolds{_p}$}
& \multicolumn{1}{c}{\textbf{Effect of confounder}} & \multicolumn{1}{c}{\textbf{Cost ratio}} &
\multicolumn{1}{c@{}}{$\bolds{95\%}$ \textbf{CI}} \\
\hline
1.1 & 2 & 1.05 & 0.88 & (0.80, 0.97) \\
1.25 & 2 & 1.05 & 0.89 & (0.81, 0.98) \\
1.5 & 2 & 1.05 & 0.91 & \textbf{(0.83, 1.01)} \\
[4pt]
1.1 & 3 & 1.05 & 0.89 & (0.81, 0.98) \\
1.25 & 3 & 1.05 & 0.91 & \textbf{(0.82, 1.00)} \\
1.5 & 3 & 1.05 & 0.94 & \textbf{(0.85, 1.03)} \\
[4pt]
1.1 & 2 & 1.1 & 0.89 & (0.81, 0.98) \\
1.25 & 2 & 1.1 & 0.92 & \textbf{(0.83, 1.01)} \\
1.5 & 2 & 1.1 & 0.96 & \textbf{(0.87, 1.06)} \\
[4pt]
1.1 & 3 & 1.1 & 0.90 & (0.82, 0.99) \\
1.25 & 3 & 1.1 & 0.95 & \textbf{(0.86, 1.04)} \\
1.5 & 3 & 1.1 & 1.02 & \textbf{(0.92, 1.12)} \\
\hline
\end{tabular*}
\legend{Unadjusted cost $\mbox{ratio} = 0.87$ (95$\%$ CI 0.79--0.96).\\
\textbf{Boldface} confidence intervals denote changes
from the unadjusted inference.\\
Mean$_0 = \mbox{mean}$ of unmeasured confounder in
cystectomy group.\\
Mean$_1 = \mbox{mean}$ of unmeasured confounder in
rad/chemo group.\\
Mean$_p = \mbox{mean}$ pooled across both groups.}
\end{table}

Income distributions are heavy-tailed. According to the U.S. census
[\citet{Census}], the ratio of the variance to the mean ($\theta$
in the
Gamma distribution) is approximately $3$ for households with members 65
and older. We took income to be measured as a ratio of individual
income to mean income. Therefore, if the true effect of income was
1.25, an increase from the mean to twice the mean would cause a 25\%
increase in medical spending. We further assumed equal $\theta$ across
treatment strata. Our sensitivity analysis (Table~\ref{tabl5}) revealed that the
treatment effect on cost became nonsignificant when income was 25\%
higher in the cystectomy group than the rad/chemo group if the effect
of income on spending was only $1.05$ and $\theta=3$. As such small
differences and effect sizes are plausible, we conclude that the
treatment effect is also sensitive to this Gamma confounder.\newpage

%s5 #&#
\section{Conclusion}\label{sec5}

We have developed a method to correct estimates of the treatment effect
on mean cost for a range of potential unmeasured confounders. The
method is simple to apply and readily accommodates potential binary,
count and continuous confounders. Although it assumes independence of
the measured and unmeasured confounders given treatment status,
simulations reveal that the adjustments are robust to modest departures
from this condition. Moreover, in all conditions we examined the
adjustment reduced bias compared to an unadjusted analysis.

Conditional independence is a generally untestable assumption that is
unlikely to hold exactly in applications. Investigators should
therefore evaluate patterns of correlation within their data before
applying the method. We recommend exploring the possible size of
partial correlations by successively omitting individual observed
confounders and estimating their partial correlations with the
remaining measured confounders. Simulations show that the adjustment is
robust to within-stratum correlations less than $0.15$, a condition
that appears to be satisfied in our bladder cancer example.

In some types of studies one expects to find substantial correlations
among measured and unmeasured confounders. For example, in a study to
examine cost of an intervention to improve student performance on a
standardized exam, measured confounders related to socioeconomic status
could be highly correlated with unmeasured confounders such as parental
attitudes toward education. If most measured confounders are tightly
interrelated with the unmeasured variable, this would render our
adjustments inaccurate. Here the investigator should use the method of
\citet{Vanderweele2011}, perhaps with simplifying assumptions
about the
confounder to render the procedure more tractable.

A second limitation of our method is that adjustments for strong
Poisson and Gamma confounders can be biased in moderate sample sizes.
The correction may also leave bias when there is a large proportion of
censored data, as would occur in studies with short follow-up. This is
likely due to the decreased efficiency of the Lin estimator when the
proportion of censoring is high, a problem that could be ameliorated by
splitting the time period into intervals [\citet{Lin2003}] or the
creation of a more efficient GLM for censored costs.

With our simple additive corrections, assuming fixed values of the
sensitivity parameters leads us to use the unadjusted $\operatorname{Var}(\hat{\beta}
^*)$ to develop confidence intervals. A fully Bayesian approach would
quantify uncertainty about the parameters using prior distributions and
propagate the uncertainty using Bayes's theorem. One could also conduct
a probabilistic sensitivity analysis using Monte Carlo simulations as
demonstrated in \citet{Arah2008}. Although it is appealing to
incorporate uncertainty about the sensitivity parameters, both methods
require sufficient knowledge about unmeasured parameters to fully
characterize their distributions.

Despite its limitations, our method provides a practical approach to
evaluate the sensitivity of costs to unmeasured confounding. The
bladder cancer study provides an example where our method can be easily
implemented to evaluate the effect of a wide range of confounders.
Further research should develop extensions to allow sensitivity
analysis for combined cost-effectiveness outcomes.

\section*{Acknowledgments}

This study used the linked SEER-Medicare database. The interpretation
and reporting of these data are the sole responsibility of the authors.
The authors acknowledge the efforts of the Applied Research Program,
NCI; the Office of Research, Development and Information, CMS;
Information Management Services (IMS), Inc.; and the Surveillance,
Epidemiology, and End Results (SEER) Program tumor registries in the
creation of the SEER-Medicare database. We thank the anonymous
reviewers and Editors for many helpful comments.

%suskaldyti doi

% imsref loaded by lrinkeviciute, 2013-08-28 09:41:03
% imsref loaded by lrinkeviciute, 2013-08-28 09:52:25
%

\printaddresses

\end{document}